\documentclass{article}

\usepackage{PRIMEarxiv}

\usepackage[utf8]{inputenc} 
\usepackage[T1]{fontenc}    
\usepackage{hyperref}       
\usepackage{url}            
\usepackage{booktabs}       
\usepackage{amsfonts}       
\usepackage{nicefrac}       
\usepackage{microtype}      
\usepackage{lipsum}
\usepackage{fancyhdr}       
\usepackage{graphicx}       
\graphicspath{{media/}}     
\usepackage{siunitx}

\newcommand{\beginsupplement}{%
        \setcounter{table}{0}
        \renewcommand{\thetable}{S\arabic{table}}%
        \setcounter{figure}{0}
        \renewcommand{\thefigure}{S\arabic{figure}}%
     }

\title{Pneumatic Computers for Embedded Control of Microfluidics
\thanks{\textit{\underline{Citation}}: 
\textbf{Ahrar S., et al., Pneumatic Computers for Embedded Control of Microfluidics,  DOI:000000/11111.}} 
}

\author
{Siavash Ahrar,$^{1,2}$ Manasi Raje,$^{1}$ 
Irene C. Lee,$^{1}$ Elliot E. Hui$^{1}$\dag\\
\\
\normalsize{$^{1}$Department of Biomedical Engineering, University of California, Irvine, USA}
\\
\normalsize{$^{2}$Department of Biomedical Engineering, CSU Long Beach, USA}\\
\\
\normalsize{\dag To whom correspondence should be addressed}
}

\begin{document}
\maketitle

\begin{abstract}
Alternative computing approaches that interface readily with physical systems are well suited for embedded control of those systems. We demonstrate finite state machines implemented as pneumatic circuits of microfluidic valves, and we employ these controllers to direct microfluidic liquid handling procedures such as serial dilution on the same chip. These monolithic integrated systems require only power to be supplied externally, in the form of a vacuum source. User input can be provided directly to the chip by covering pneumatic ports with a finger. State machines with up to four bits of state memory are demonstrated, and next-state combinational logic can be fully reprogrammed by changing the hole-punch pattern on a membrane in the chip. These pneumatic computers demonstrate a new framework for the embedded control of physical systems.
\end{abstract}


\section{Introduction}
Embedded digital controllers are widely employed for directing physical operations in integrated systems. Transduction between the electronic and physical realms can sometimes be unwieldy however, requiring substantial mechanical hardware that can potentially dominate a system. It would thus be attractive to perform computing in alternative media that are more similar to the physical systems requiring control. This goal is being actively pursued in a variety of fields ranging from robotics \cite{Wehner2016,Preston02} to biology \cite{Elowitz1183,Bonnet1,Roquetaad8559}, but such control systems remain rudimentary compared to their electronic counterparts. Microfluidics has produced powerful biochemical tools \cite{Jacobs1,Ottesen1,Macosko1} and also demonstrated strong potential for digital computing \cite{Jensen01,Prakash832,Mosadegh02,Duncan01,Frontier37}, making it a promising platform for the development of non-electronic embedded control.

\section{Results}

The finite state machine (FSM) is a widely employed abstraction in digital control systems and can be implemented efficiently with a limited number of logic gates. We set out to build FSM controllers out of microfluidic circuits and to employ these controllers to direct liquid handling processes on the same microfluidic chip. Boolean logic was implemented by using circuits of pressure-actuated microfluidic membrane valves \cite{Jensen01, Duncan01,Frontier37,Rhee01,Weaver01,Devaraju01}. Specifically, we employed gates analogous to N-type metal-oxide-semiconductor (NMOS) logic, with normally closed membrane valves \cite{Grover02} forming the pull-down network and long microfluidic channels providing pull-up resistance. Vacuum pressure represents binary 1, and atmospheric pressure represents binary 0. We have previously shown that this implementation achieves excellent pressure gain \cite{Duncan01,Duncan02} , which is critical for achieving robustness to noise and error-free data propagation through cascaded gates with fan-out.

An FSM consists of a set of program states, each with associated functional outputs, and a set of rules governing the transitions between different states. In a digital implementation, the system state is stored as a binary code in the state register, state transitions are calculated by combinational Boolean logic, and a system clock controls the timing of the transitions. We implemented state registers by employing negative-edge-triggered D flip-flops \cite{Rhee01,Devaraju01} in a clocked leader-follower configuration to render the system robust against race conditions. The output of the state register (representing the current system state) is input into a combinational logic block along with any system inputs to calculate the next system state. During the subsequent clock cycle, the next state becomes the current state.

We began with a demonstration of conditional branching in a simple FSM with two bits of state memory and one input. This FSM program (state transition diagram shown in Fig. 1A) has two loops, one with the two bits oscillating in phase (00 or 11) and the other with the two bits oscillating out of phase (01 or 10). The FSM remains in the current loop when the input is 1 and toggles to the other loop when the input is 0. The next state logic for this program can be implemented with one NOT gate and one NAND gate. In fact, the entire FSM was implemented with only 16 valves (Fig. 1B). An additional two valves were utilized as visual indicators of the current state. Circuit operation was filmed by adjusting the lighting to reflect differently from open and closed valves and then processed for presentation as a circuit timing diagram (Fig. 1C).

Next, we demonstrated embedded control by integrating an FSM with a rotary mixer \cite{Chou01,Nguyen01}. Here, the FSM controls valve settings in the mixer to direct a sequence of liquid handling operations (Fig. 2A). Liquid flow is driven by a peristaltic pump controlled by an on-chip ring oscillator \cite{Duncan01}. The FSM was programmed to loop sequentially through four states. In the first state (10), the entire ring is filled from Reservoir 1 (Fig. 2A). In the second state (11), the left half of the ring is filled from Reservoir 2. In the third state (01), the entire ring is filled from Reservoir 2. In the fourth state (00), both reservoirs are sealed off and the contents of the ring are circulated for mixing. The mixing loop can be loaded with the contents of the two reservoirs at a variety of ratios, depending on whether the third state (01) is allowed to run to completion or is terminated early. This functionality is illustrated in Fig. 2D where in the first instance the loop is yellow during the mixing step (State 00), and in the second instance the loop is a mix of blue and yellow.

Since the composition of the loop during the mixing step (State 00) is determined by the timing of the state transitions, we engineered an input method for the user to trigger these transitions. The system clock was directly connected to the vacuum supply, but the connection was shorted to ground by a port that is open to the atmosphere (Fig. 2A). While the port is uncovered, the clock signal holds steady at 0, but when the port is covered, the clock signal switches to 1. Thus, the user can advance the FSM to the next state simply by covering and then releasing the port with a finger. Please refer to Movie S1. Notably, aside from the need for a vacuum supply, the entire system is self-contained on a single microfluidic chip (Fig. 2B). Liquids are pipetted directly onto the chip, and all timing signals are generated by the on-chip pneumatic oscillator or by user input to the on-chip button. This monolithic integrated system was built by using 31 valves, 17 for the FSM controller, and 14 for the rotary mixer.

As a more complex example of embedded control, we chose to implement serial dilution, a ubiquitous multistep liquid-handling procedure for creating a logarithmic range of solution concentrations. We built upon our previously reported dilution ladder design \cite{Ahrar01}, which accurately performs a sequence of 1:1 dilutions and preserves the entire dilution series. While the earlier work required off-chip solenoid valves under computer control, here we employed an on-chip FSM for embedded control. The design consists of a set of liquid compartments arranged as rungs on a ladder. For each dilution step, two neighboring rungs are connected in a loop and liquid is circulated until mixing is complete (Fig. 3C). Timing signals for peristaltic pumping were generated by an on-chip ring oscillator and routed by the FSM to the appropriate rungs of the ladder for each stage of dilution (Movie S2). A 4-bit FSM was employed to drive the four control lines of the ladder directly via one-hot encoding, thus reducing chip real estate by avoiding the need for a demultiplexer. The autonomous serial dilution system is the largest circuit presented in this work (Fig. 3B), employing 81 valves: 38 valves for the FSM, 15 for the dilution ladder, 5 for the ring oscillator, and 23 valves for buffering and routing signals from the oscillator to the dilution ladder.

Finite state machines are defined by the rules that govern the transitions between states. Programmability thus entails the ability to specify these rules, which we achieved by employing a programmable logic array (PLA) to implement next-state logic. In the FSM design process, a state diagram is converted into a state transition table, which is then converted into a set of sum-of-products Boolean logic expressions (Fig. S1). These excitation equations are encoded into the microfluidic PLA by boring a set of holes in the membrane layer of the chip to define interconnects between the two circuit layers of the device. The PLA employs a NAND-NAND architecture with six inputs (2 state bits and 1 input bit, plus their inverses) and two outputs (2 next-state bits). The PLA design was initially verified as a standalone device (Figs. S2 and S3) before being integrated with state registers (Fig. 4A) to enable complete FSMs to be formed. The system requires 35 valves: 18 for the PLA, 13 for state memory, and 4 valves for signal buffering. An additional 4 valves were employed as signal indicators. We implemented three different FSM programs sequentially on a single physical chip. Reprogramming was accomplished by exchanging the membrane layer to obtain different interconnect hole patterns in the PLA circuit. The system was operated successfully at clock frequencies up to 5 Hz (Movies S3 and S4). Circuit operation data for each FSM program are presented as timing diagrams in Fig. 4B-4D.	 

\section{Discussion}
Lab-on-a-chip systems generally require substantial off-chip mechanical and electronic components to operate. While simple event staging can be achieved in standalone chips based on capillary flow \cite{Toley1}, programmable operation of integrated valve networks requires control systems that greatly exceed both the cost and size of their associated microfluidic devices \cite{Brower1}. On-chip microfluidic logic circuitry has already been deployed in commercial products in order to realize complex microfluidic functionality that otherwise would require an unmanageable number of pneumatic connections between the chip and controller \cite{Frontier37}. Here, we take the next step to demonstrate a fully embedded control strategy that allows off-chip control to be dispensed with entirely. This approach produces monolithic systems-on-a-chip amenable to low-cost batch fabrication and exhibiting highly compact form factors, greatly reducing the barriers to broad distribution of lab-on-a-chip systems. Pneumatic FSMs are also well suited to bring powerful new capabilities to the embedded control of soft robotics, where pneumatic circuitry is already attracting substantial interest \cite{Wehner2016,Preston7750,Drotman1,Hoang1}. Adoption of pneumatic digital circuitry can be catalyzed by the development of rapid prototyping approaches \cite{Werner01} and design automation tools \cite{Sanka01} to accelerate development cycles.

Integrated circuits that can be configured by customers after manufacturing are important commercially, particularly in embedded systems. The microfluidic systems presented in this work consist of two glass layers containing etched channels, sandwiched around a flat membrane of silicone elastomer. For bulk manufacturing, the glass layers could be replaced by injection-molded plastic, and membrane through-holes could be defined by laser cutting \cite{Duncan02}. A single injection-molded design could thus be assembled with a number of different laser-cut membranes to implement a variety of different assays, each driven by a different control program. Programmability thus provides a strategy to produce a large assortment of niche products at a cost per device comparable to a mass-market product, through amortizing the cost of mold production across a set of dissimilar products. While vacuum supply is not widely available outside of laboratories and hospital rooms, the systems presented in this work can be powered by the vacuum from a small electrical pump, or even a manual pump or syringe \cite{Duncan01}. Thus, this technology can support portable, standalone, biochemical processing systems.

A variety of non-electronic computing approaches have been demonstrated at the proof-of-concept level, with the FLODAC fluidic microprocessor representing one of the most advanced efforts \cite{Gluskin}. However, information processing via physical systems faces inherent speed and density disadvantages compared to electronics, and systems like FLODAC simply could not compete with microelectronics. On the other hand, physical computing has distinct advantages for controlling physical systems, as transducers become unnecessary when control signals are calculated in the same physical modality as the system that requires control. Indeed it was in the control of air-conditioning units, lawn sprinklers, and showerheads that fluidics was able to find commercial success \cite{Joyce1}. Upon this foundation, we now add a programmable digital control architecture miniaturized for the control of contemporary applications such as microfluidics and soft robotics.

\newpage
\section* {Supplementary material}

\textbf {Fabrication}: Devices were fabricated as previously described \cite{Duncan01}. Briefly, microchannels were defined in glass by photolithography and wet etching (50 µm deep). Ports were drilled through the glass to provide channel access. The channels were covered by a silicone sheet (254 µm) and vias were manually punched through the silicone under microscope observation. A second layer of glass channels was then laid on top of the silicone, again using a microscope to ensure proper alignment, and the layers were pressed together manually. 

\textbf{Device operation}: Switching of pneumatic inputs was accomplished via computer-controlled solenoid valves. Clock signals were often generated off chip to provide greater flexibility and control during circuit characterization. On-chip timing control by using oscillator circuits or manual push buttons was also employed.

\textbf {Circuit analysis}: Device operation was filmed with a DSLR camera (Canon EOS Rebel T1, 200 mm lens). Chips were positioned to maximize the difference in light reflection between open and closed valves. Circuit timing diagrams were extracted in MATLAB by plotting the normalized pixel intensity of selected regions of interest.

\textbf {Programmable Logic Array (PLA)}: The PLA is a programmable combinational logic circuit designed to allow a large variety of Boolean sum-of-products functions to be encoded. A PLA is typically organized as a set of AND gates feeding into a set of OR gates. We chose to implement our PLA using the DeMorgan equivalent of NAND gates feeding into NAND gates, since NAND gates are more straightforward to implement in our technology.

To implement a particular FSM design, the state transition diagram is translated into a truth table mapping each combination of states and inputs to a specific next state. This can then be translated into sum-of-products Boolean logic functions defining each next-state bit as a function of the state bits and input bits. In order to encode these next-state logic functions into our PLA implementation, bore holes are drilled in the membrane layer separating the two channel layers of the device. In this way, specific PLA inputs are connected to specific inputs of the initial AND gates. Similarly, bore holes are drilled to connect specific AND gate outputs to specific OR gate inputs. Defining these AND functions and OR functions implements the sum-of-products Boolean functions representing the desired FSM program.

A standalone PLA system is demonstrated in Figs. S1 and S2. This system includes indicator valves that allow the state of each input and output to be monitored simultaneously by imaging the chip. By imaging the inputs and outputs for all possible input combinations, visual truth tables were constructed to verify accurate implementation of two different sets of Boolean functions. The same device is used to implement two different sets of Boolean logic, with the only difference being a change in the membrane bore hole patterns.

\textbf {List of Supplementary Videos}:
\begin{itemize}
  
  \item Movies S1 Embedded control of microfluidic liquid handling 
  \item Movies S2 Autonomous control of serial dilution 
  \item Movies S3 Programmable FSM First Encoding (clock frequency 1 Hz)
 \item  Movies S4 Programmable FSM third encoding (clock frequency 5 Hz)
\end{itemize}

\newpage

\section* {Acknowledgments}  
Authors thank DJ Li, A Ravikumar, B Wong, and EM Werner for helpful feedback on the manuscript. Authors express gratitude to Hui-Lab members PN Duncan, PV Thomas, and TV Nguyen. Additionally, authors acknowledge WH Grover for his generous support at the start of the project and manuscript feedback.\\ 
This work was supported in part by the NSF (ECCS-1102397) and the DARPA N/MEMS Fundamentals Program under N66001-1-4003 issued by SPAWAR to the Micro/Nano Fluidics Fundamentals Focus Center. SA. was also supported in part via the NSF LifeChips IGERT (0549479) fellowship. Research reported in this publication was also supported by the National Institute Of General Medical Sciences of the National Institutes of Health under Award Number R01GM134418. The content is solely the responsibility of the authors and does not necessarily represent the official views of the National Institutes of Health.

\section*{Affiliations}
S.A. UC Irvine, Biomedical Engineering;\\
and CSU Long Beach, Biomedical Engineering.\\
M.R. UC Irvine, Biomedical Engineering.\\
I.C.L. UC Irvine, Biomedical Engineering.\\
E.E.H. UC Irvine, Biomedical Engineering. 

\section*{Addresses}
1. Department of Biomedical Engineering, University of California, 3120 Natural Sciences II, Irvine, CA 92697-2715, USA. E-mail: eehui@uci.edu 
2. Department of Biomedical Engineering, California State University Long Beach, 1250 Bellflower Blvd.
VEC-404.A, Long Beach, CA 90840, USA. E-mail: siavash.ahrar@csulb.edu

\bibliographystyle{unsrt}  
\bibliography{references}  

\newpage

\begin{figure}[b]
\includegraphics[width=\textwidth]{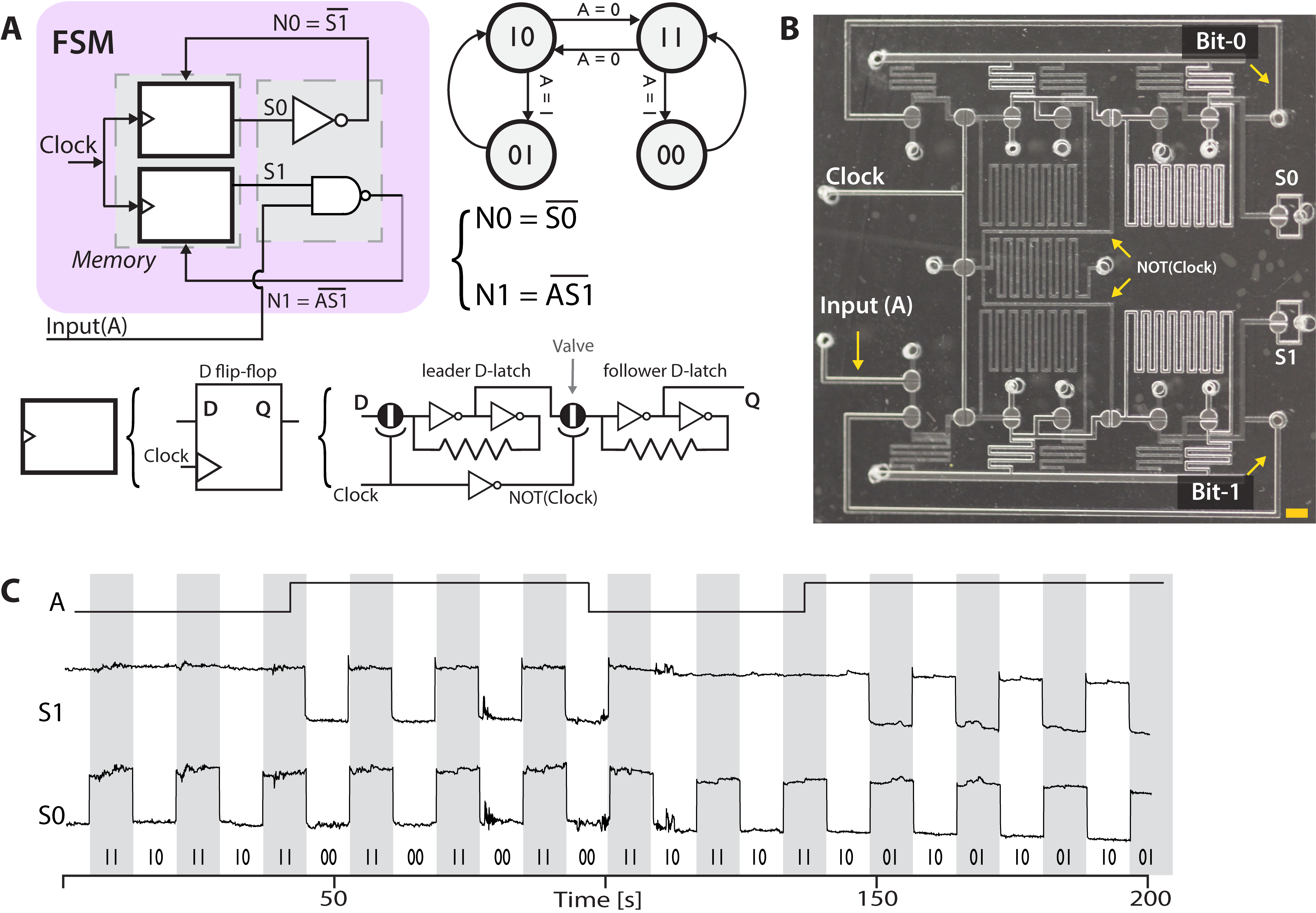}
\centering
\caption{\textbf {Pneumatic Finite State Machine} (A) 2-bit FSM block diagram and state transition diagram. Conditional branching is demonstrated by employing a single input. Pneumatic valves are 3-terminal devices in which the channel is opened by applying vacuum to the gate. (B) Annotated image of microfluidic chip. Scale bar is 1.5 mm. (C) Circuit timing diagram. System state is measured via valves S1 and S0.}
\end{figure}

\begin{figure}[h]
\includegraphics[width=\textwidth]{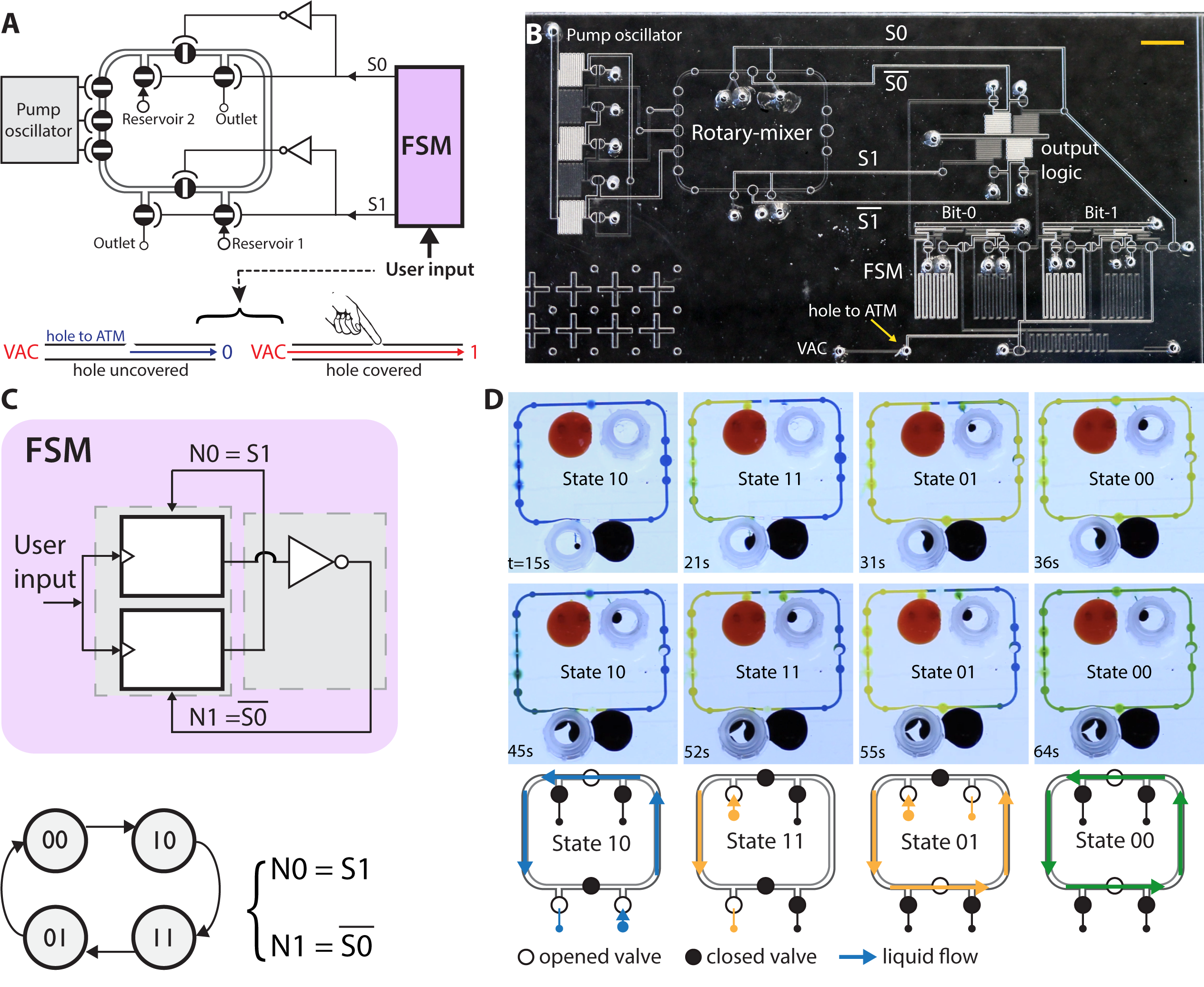}
\centering
\caption{\textbf {Embedded Control of Microfluidic Liquid Handling} (A) System-level diagram of rotary mixer with 2-bit FSM controller. Push-button input is created by shorting a vacuum input to atmospheric ground through a hole that can be covered by the user. (B) Annotated image of microfluidic chip. Scale bar is 5 mm. (C) FSM block diagram and state transition diagram. (D) Images of the rotary mixer progressing through different states. The duration of State 01 determines the composition of the mixer in State 00 and can be controlled by the user by tapping on the push-button input.}
\end{figure}

\begin{figure}[h]
\includegraphics[width=\textwidth]{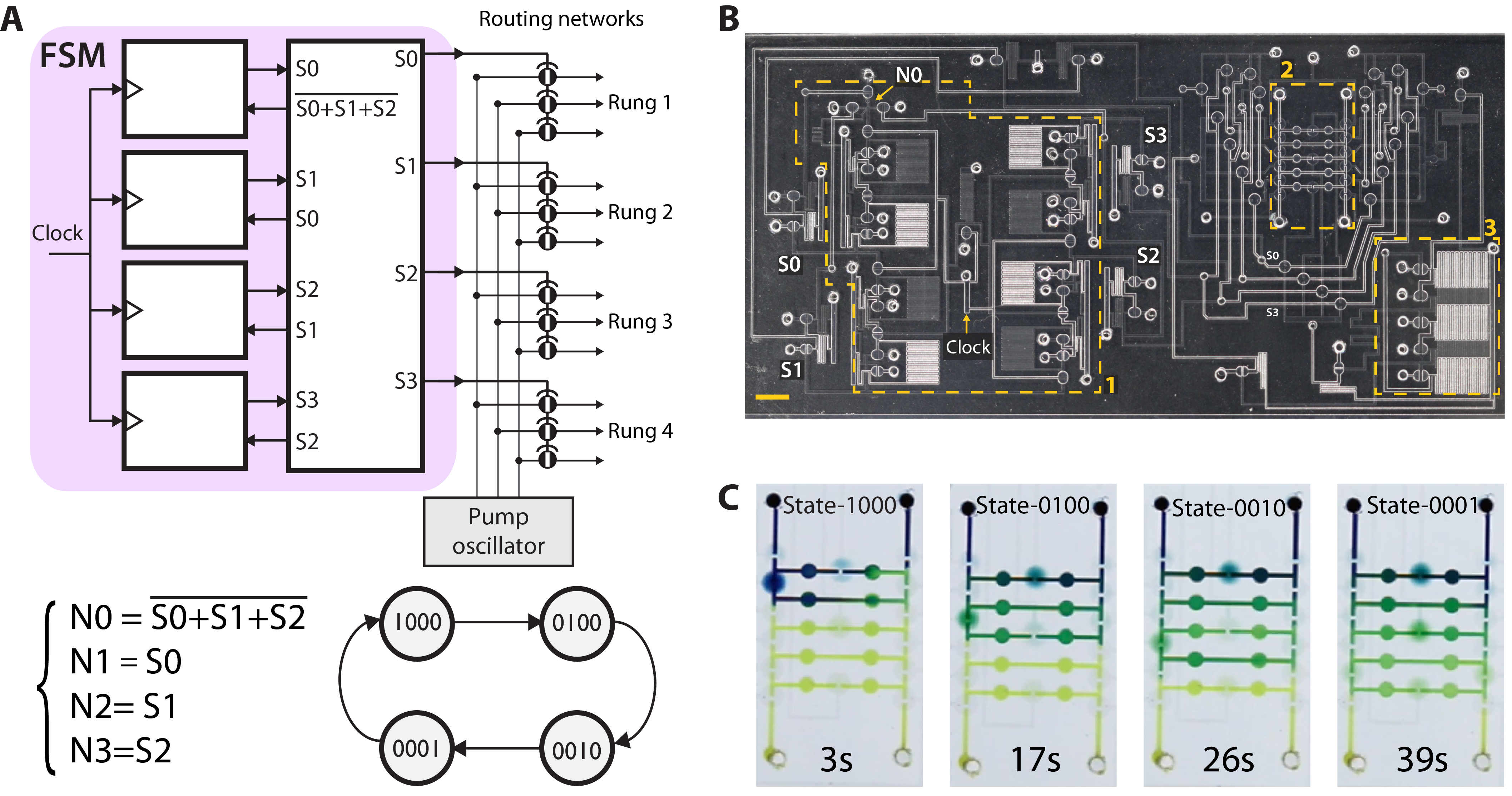}
\centering
\caption{\textbf {Autonomous Control of Serial Dilution} (A) System-level diagram of the autonomous dilution ladder. A 4-bit FSM controls routing of the pump oscillator to the appropriate rungs of the dilution ladder. (B) Annotated image of microfluidic chip: FSM (Box 1), dilution ladder (Box 2), ring oscillator (Box 3), and the pump control routing network making up the remainder of the chip. Scale bar is 5 mm. (C) Images of 1:1 serial dilution process.}
\end{figure}

\begin{figure}[h]
\includegraphics[width=0.9\textwidth]{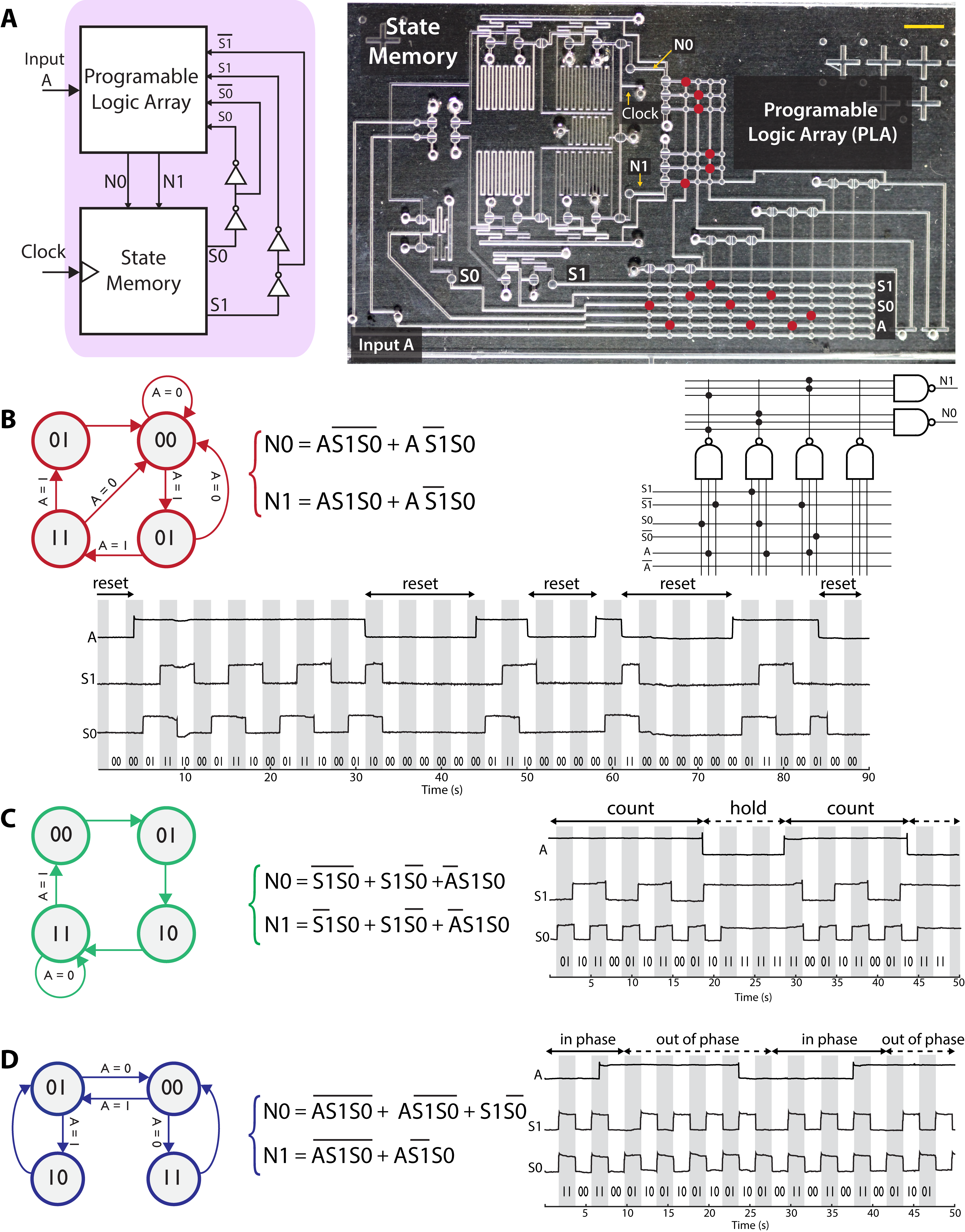}
\centering
\caption{\textbf {Programmable FSM Encoded Three Ways}: (A) Block diagram and annotated image of microfluidic chip. Red circles represent location of membrane bore holes to implement FSM of panel B. Scale bar is 5 mm. (B) FSM steps through each state sequentially and resets to state 00 whenever A=0. (C) FSM steps through each state sequentially and holds at 11 when A=0. (D) FSM states oscillate in phase when A=0 and out of phase when A=1.}
  \label{figurelabel}
\end{figure}

\beginsupplement

\newpage
\begin{figure}[t]
\includegraphics[width=\textwidth]{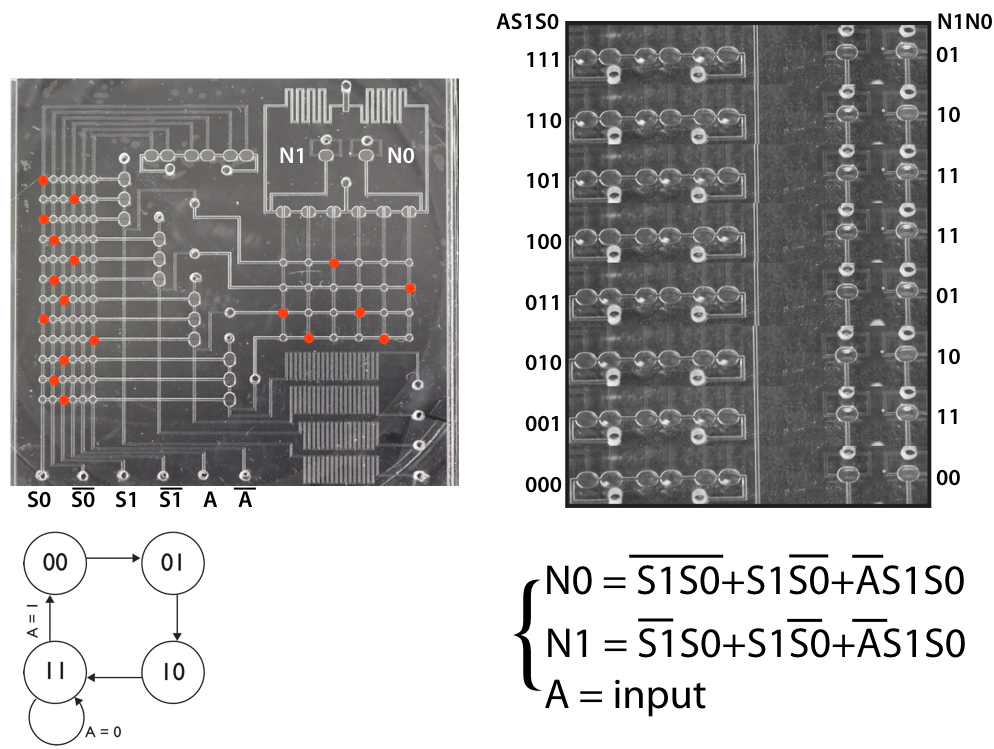}
\centering
\caption{\textbf {Programmable Logic Array, Program 1}: 
Programmable combinational logic with 6 inputs and 2 outputs. Inputs represent the current state (S0, S1) and one input (A), along with the inverses of these three variables. Outputs represent the next state (N0, N1). The PLA is organized as four 3-input NAND gates followed by two 3-input NAND gates, which is equivalent to four AND gates followed by two OR gates. Bore holes through the membrane layer (indicated by the red dots) allow connections between horizontal channels on the first circuit layer and vertical channels on the second circuit layer. The bore hole positions were chosen to encode the set of Boolean functions shown, defining each next-state bit as a function of the current state and the system input. These sum-of-products excitation equations represent the FSM program summarized in the state transition diagram shown. Outputs were measured for all 8 possible inputs, generating a visual truth table that verifies faithful implementation of the Boolean functions.}
\end{figure}

\newpage
\begin{figure}[t]
\includegraphics[width=\textwidth]{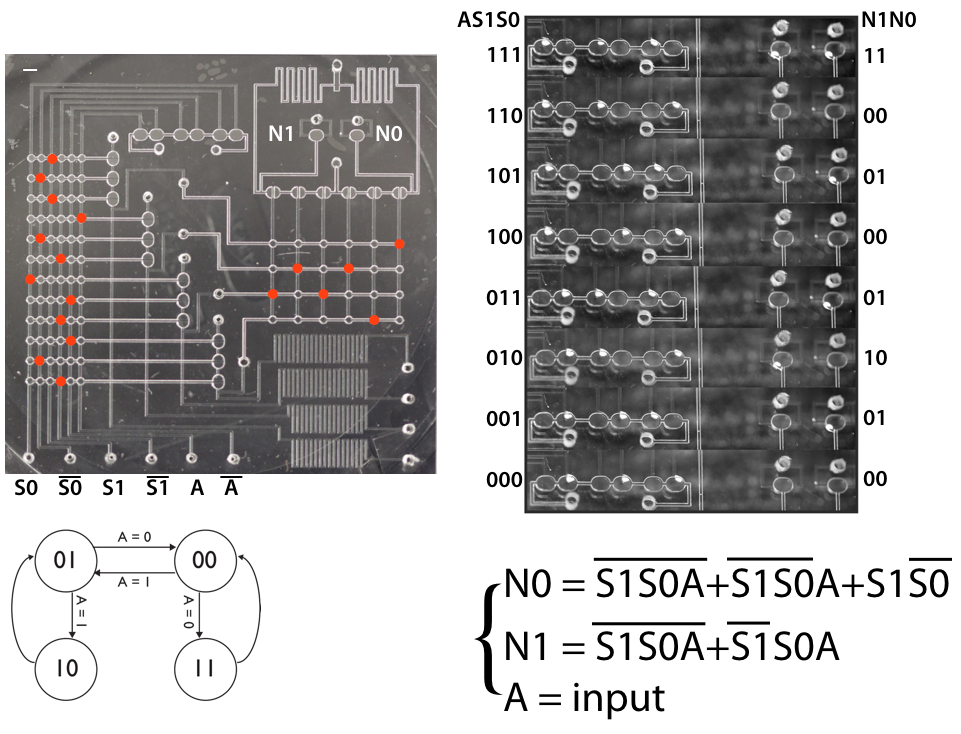}
\centering

\caption{\textbf {Programmable Logic Array, Program 2}: 
The device of Figure S1, but here it includes a different bore hole pattern (red dots) to encode a different set of Boolean functions,  representing a different FSM state transition diagram. Outputs were measured for all 8 possible inputs, generating a visual truth table that verifies faithful implementation of the Boolean functions.}
\end{figure}

\end{document}